\documentclass[3p,times]{elsarticle}

\usepackage{fancyvrb}
 \biboptions{compress}
\usepackage{color}

\usepackage[figuresright]{rotating}
\usepackage{amssymb}
\newfont{\fp}{msbm10 at 11pt}  

\usepackage{color, colortbl, framed}
\usepackage{amsmath,amsfonts,graphicx,color}

\usepackage{amsfonts}

\begin{document}

\begin{frontmatter}

\title{Statistical test for fractional Brownian motion based on detrending moving average algorithm}


\author[label1]{Grzegorz Sikora}
\address[label1]{Faculty of Pure and Applied Mathematics, Hugo Steinhaus Center\\     
    Wroclaw University of Science and Technology, Janiszewskiego 14a, 50-370 Wroc{\l}aw, Poland}
\ead{grzegorz.sikora@pwr.edu.pl}
\begin{abstract}
Motivated by contemporary and rich applications of anomalous diffusion processes we propose a new statistical test
for fractional Brownian motion, which is one of the most popular models for anomalous diffusion systems. The test is
based on detrending moving average statistic and its probability distribution. Using the theory of Gaussian quadratic
forms we determined it as a generalized chi-squared distribution. The proposed test could be generalized for statistical
testing of any centered non-degenerate Gaussian process. Finally, we examine the test via Monte Carlo simulations
for two exemplary scenarios of subdiffusive and superdiffusive dynamics.
\end{abstract}

\end{frontmatter}

\section{Introduction}
The theory of stochastic processes is currently an important and developed branch of mathematics \cite{Bor17,CoxMil17,Dur16}. The key issue from the point of view of the application of stochastic processes is statistical inference for such random objects \cite{Lin04,MisShe17,Raj13,Rao14}. This field consists of statistical methods for the reliable estimation, identification, and validation of stochastic models. Such a part of the theory of stochastic processes and the statistics developed for them are used to model phenomena studied by other fields such as physics \cite{FrePos10,PauBas13,Kam07}, chemistry \cite{FrePos10,Sch16,Kam07}, biology \cite{Bre14, CapBak15,FrePos10,Hol17,Sch14,Sch16}, engineering \cite{Bei06,Sco12}, among others.

This work is motivated by growing interest and applications of the special class of stochastic processes, namely anomalous diffusion processes, which largely depart from the classical Brownian diffusion theory \cite{MetKla00, Rak14}. Such processes are characterized by a nonlinear power-law growth of the mean squared displacement (MSD) in the course of time. Their anomalous diffusion behavior manifested by nonlinear MSD is intimately connected with the breakdown of the central limit theorem, caused by either broad distributions or long-range correlations. Today, the list of systems displaying anomalous dynamics is quite extensive \cite{EfrNes16,GudDyb10,KepBro11,MagKla10, NegIna17, Pal11}. Therefore in recent years, there has been great progress in the understanding of the different mathematical models that can lead to anomalous diffusion \cite{KlaLim12, KlaSok11, MerSok15}. One of the most popular of them is the fractional Brownian motion (FBM) \cite{Ful17,JeoMet10,KepBro11,LisTot18,MerSok15,SzyWei09,Wei13}. Introduced by Kolmogorov \cite{Kol40} and studied by Mandelbrot in a series of papers \cite{Man83,ManNes68}, it is now well-researched stochastic process. FBM is still constantly developed by mathematicians in different aspects \cite{BahTor18,DavHas09,Muk17,Nou12,WanChe18}. 

The main subject considered in this work is the issue of rigorous and valid identification of the FBM model. The problem of FBM identification has been described in the mathematical literature for a long time \cite{Ber94,Coe00}. However, most of the works mainly concern various methods of estimating the parameters of the FBM model. They are based, among others, on p-variation \cite{MagWój13}, discrete variation \cite{Coe01}, sample quantiles \cite{Coe08} and other methods \cite{Bon12,BreCoe12,CoeKor12,Haj15, LiuLiu09, MieWoj07,TaqTev95, Yer14}. A certain gap in this theory is the lack of tools such as rigorous statistical tests to identify the FBM model in empirical data. Some approaches to FBM identification are known, e.g., application of empirical quantiles \cite{BurKep12}, distinguishing FBM from pure Brownian motion \cite{Li16}. 
According to the author's current knowledge, the only statistical test for the FBM model is the test based on the distribution of the time average MSD \cite{SikBur17}. Due to the lack of statistical tests specially designed for the FBM model, in this work, we propose such a statistical testing procedure. 

The proposed test has a test statistic which is the detrending moving average (DMA) statistic introduced in the paper \cite{AleCar02}. For more than a decade, the DMA algorithm has become an important and promising tool for the analysis of stochastic signals. It is constantly developed and improved \cite{AriCar17,Car09,CarKiy16,Sha15}, its multifractal version was created and used \cite{Car07,Gu10, Jia11,  Xio14} and it is applied for different empirical datasets \cite{LiCao18,PalRao14, PonCar17,Ser10}. As one of the important method for fluctuation analysis, the DMA algorithm was often compared with other methods \cite{BasBar08,XiZha16,SinChe11} 

In section \ref{sec:dis} we show that the distribution of the DMA statistics follows the generalized chi-squared distribution. The main section \ref{sec:test} demonstrates the statistical testing procedure based on computing the DMA statistic for empirical data. In section \ref{sec:sim} the results of Monte Carlo simulations of the proposed test are presented and discussed. Section \ref{sec:con} contains conclusions and final remarks. In the last section \ref{sec:app}, the Matlab code of the proposed test is presented. 

\section{Probability distribution of DMA statistic}\label{sec:dis}
The DMA algorithm was introduced in \cite{AleCar02}. For a finite trajectory $\{X(1),X(2),\ldots,X(N)\}$ of a stochastic process the DMA statistic has the following form
\begin{equation}\label{for1}
\sigma^2(n)=\frac{1}{N-n}\sum_{j=n}^{N}(X(j)-\tilde{X}_n(j))^2,\quad n=2,3,\ldots,N-1,
\end{equation}
where $\tilde{X}_n(j)$ is a moving average of $n$ observations $X(j),\ldots,X(j-n+1)$, i.e.
$$\tilde{X}_n(j)=\frac{1}{n}\sum_{k=0}^{n-1}X(j-k).$$
The statistic $\sigma^2(n)$ is a random variable which computes the mean squared distance between the process $X(j)$ and its moving average $\tilde{X}_n(j)$ of the window size $n.$ It has scaling law behavior $\sigma^2(n)\sim C_Hn^{2H},$ where $H$ is a self--similarity parameter of the signal \cite{AleCar02,AriCar17}. The constant $C_H$ has explicit expression computed in the case of fractional Brownian motion \cite{AriCar17}. As a byproduct of this scaling law one can estimate the self--similarity parameter $H$ from linear fitting on double logarithmic scale \cite{BasBar08,CarKiy16,ShaGu12}.

In this work, we leave the issue of DMA algorithm as an estimation method and concentrate on the probability characteristics of this random statistic.  Throughout the paper, we assume that the stochastic process $X(j)$ is a centered Gaussian process. Therefore a finite trajectory $\mathbb{X}=\{X(1),X(2),\ldots,X(N)\}$ is a centered Gaussian vector with covariance matrix $\Sigma=\{E\left[X(j)X(k)\right]:j,k=1,2,\ldots,N\}.$ Let introduce the process $Y(j):=X(j+n-1)-\tilde{X}_n(j+n-1),$ which is still a centered Gaussian process. We calculate the covariance matrix of the vector $\mathbb{Y}=\{Y(1),Y(2),\ldots,Y(N-n+1)\}$ 
\begin{eqnarray}\nonumber
E\left[Y(j)Y(k)\right]&=&E\left[X(j+n-1)X(k+n-1)\right]-E\left[X(j+n-1)\tilde{X}_n(k+n-1)\right]-E\left[\tilde{X}_n(j+n-1)X(k+n-1)\right]\\\nonumber
&+&E\left[\tilde{X}_n(j+n-1)\tilde{X}_n(k+n-1)\right]=E\left[X(j+n-1)X(k+n-1)\right]-\frac{1}{n}\sum_{m=k}^{k+n-1}E\left[X(j+n-1)X(m)\right]\\
&-&\frac{1}{n}\sum_{l=j}^{j+n-1}E\left[X(k+n-1)X(l)\right]+\frac{1}{n^2}\sum_{j\leq l\leq j+n-1}\sum_{k\leq m\leq k+n-1}E\left[X(l)X(m)\right].\label{for2}
\end{eqnarray}
That matrix we denote by $\tilde{\Sigma}=\{E\left[Y(j)Y(k)\right]:j,k=1,2,\ldots,N-n+1\}.$
We see that the dependence structure of the process $Y(i)$ is fully determined by the covariance of the process $X(i).$ Moreover the covariance $E\left[X(k)X(m)\right]$ in formula \eqref{for2} has a prefactor 
\begin{eqnarray*}
\left(1-\frac{1}{n}\right)^2,&\textrm{ for }&l=j+n-1 \wedge m=k+n-1,\\
\frac{1}{n^2}-\frac{1}{n},&\textrm{ for }&(l=j+n-1\wedge m\neq k+n-1)\vee(l\neq j+n-1\wedge m=k+n-1),\\
\frac{1}{n^2},&\textrm{ for }&l\neq j+n-1 \wedge m\neq k+n-1.
\end{eqnarray*}
Therefore we can rewrite the formula \eqref{for2} in the equivalent form
\begin{eqnarray}
E\left[Y(j)Y(k)\right]&=&\left(1-\frac{1}{n}\right)^2E\left[X(j+n-1)X(k+n-1)\right]+\left(\frac{1}{n^2}-\frac{1}{n}\right)\left[\sum_{m=k}^{k+n-2}E\left[X(j+n-1)X(m)\right]\right.\nonumber\\&+&\left.\sum_{l=j}^{j+n-2}E\left[X(l)X(k+n-1)\right]\right]+\frac{1}{n^2}\sum_{j\leq l\leq j+n-2}\sum_{k\leq m\leq k+n-2}E\left[X(l)X(m)\right].\label{for3}
\end{eqnarray}
The average value of random variable $\sigma^2(n)$ we can now express based on \eqref{for2} and \eqref{for3} by covariance structure of the process $X(j)$
\small\begin{eqnarray}
\nonumber E\left[\sigma^2(n)\right]&=&\frac{1}{N-n}\sum_{j=n}^{N}E\left[\left(X(j)-\tilde{X}_n(j)\right)^2\right]=\frac{1}{N-n}\sum_{j=n}^{N}E\left[Y^2(j-n+1)\right]\\&=&\!\!\!\!\!\frac{1}{N-n}\sum_{j=n}^{N}\!\!\left\{\left(1-\frac{1}{n}\right)^2\!\!E[X^2(j)]+2\left(\frac{1}{n^2}-\frac{1}{n}\right)\!\!\sum_{m=j-n+1}^{i-1}\!\!\!\!E[X(j)X(m)]+\frac{1}{n^2}\!\!\sum_{j-n+1}^{j-1}\!E[X^2(m)]+\frac{2}{n^2}\!\!\!\sum_{j-n+1\leq k<m\leq j-1}\!\!\!\!\!\!\!\!E[X(m)X(l)]\right\}.\label{for3b}
\end{eqnarray}
We can also express the variance of the random variable $\sigma^2(n)$
\small\begin{eqnarray}\label{for3c}\nonumber
Var\left[\sigma^2(n)\right]&=&\frac{1}{(N-n)^2}Var\left[\sum_{j=n}^{N}Y^2(j-n+1)\right]=\frac{1}{(N-n)^2}\sum_{l,m=n}^{N}Cov\left(Y^2(l-n+1),Y^2(m-n+1)\right)\\
&=&\frac{1}{(N-n)^2}\sum_{l,m=n}^{N}E\left[Y^2(l-n+1)Y^2(m-n+1)\right]-E\left[Y^2(l-n+1)\right]E\left[Y^2(m-n+1)\right].
\end{eqnarray}
\normalsize
The terms $E\left[Y^2(l-n+1)\right]$ and $E\left[Y^2(m-n+1)\right]$ one can compute from covariance of the process $Y(j)$ according to \eqref{for3}.
The 4th--order moment $E\left[Y^2(l-n+1)Y^2(m-n+1)\right]$ can be expressed by covariance structure of process $Y(j)$ according to Isserlis' theorem \cite{SonLee15}:
$$E\left[Y^2(l-n+1)Y^2(m-n+1)\right]=E\left[Y^2(l-n+1)\right]E\left[Y^2(m-n+1)\right]+2E\left[Y(l-n+1)Y(m-n+1)\right]^2.$$
Therefore applying above to \eqref{for3c} we get
\begin{equation}\label{for3d}
Var\left[\sigma^2(n)\right]=\frac{2}{(N-n)^2}\sum_{l,m=n}^{N}E\left[Y(l-n+1)Y(m-n+1)\right]^2.
\end{equation}
Using \eqref{for3} one can present formula \eqref{for3d} for variance in terms of covariance structure of the underlying process $X(i).$

\normalsize
In order to describe more probabilistic properties of the random variable $\sigma^2(n)$ we notice the quadratic form representation
$$\sigma^2(n)=\frac{1}{N-n}\sum_{j=n}^{N}Y^2(j-n+1)=\frac{1}{N-n}\mathbb{Y}\mathbb{Y}^T,$$
where $\mathbb{Y}^T$ is a vertical vector which is a transpose of the vector $\mathbb{Y}.$ The random object $\mathbb{Y}\mathbb{Y}^T$ is a quadratic form of a Gaussian vector $\mathbb{Y}.$ Therefore we apply the theory of Gaussian quadratic forms to study random variable $(N-n)\sigma^2(n).$ The theory of Gaussian quadratic forms \cite{MatPro92} provides us with the following representation
\begin{equation}
(N-n)\sigma^2(n)\stackrel{d}{=}\sum_{j=1}^{N-n+1}\lambda_j(n)U_j
\label{for4},
\end{equation}
where $\stackrel{d}{=}$ means equality in distribution. The probability distribution in \eqref{for4} is the generalized chi-squared distribution \cite{dav80}. The random variables $U_j'$s form an i.i.d. sequence of chi-squared distribution with one degree of freedom. The coefficients $\lambda_j(n)$ are the eigenvalues of covariance matrix $\tilde{\Sigma}$ of the vector $\mathbb{Y}.$ They depend on $n$ and the parameters of the process $Y(j).$ The distribution in \eqref{for4} one can interpret as a sum of independent gamma distributions with constant shape parameter $1/2$ and different scale parameters, i.e. $\lambda_j(n)U_j\stackrel{d}{=}G(1/2,2\lambda_j(n)).$ By $G(\alpha,\beta)$ we denote the gamma distribution with the shape parameter $\alpha$ and scale parameter $\beta$. It has a PDF of the form
$$f_{(\alpha,\beta)}(x)=\frac{x^{\alpha-1} \,  \exp(-x/\beta)}{\Gamma(\alpha) \,\beta^\alpha} \quad (x>0)$$
and CDF
$$F_{(\alpha,\beta)}(x)=\frac{1}{\Gamma(\alpha)}\gamma(\alpha,x/\beta),$$
where $\Gamma$ function and lower incomplete gamma function $\gamma$ are defined respectively
$\Gamma(z)=\int_{0}^{\infty}x^{z-1}e^{-x}dx$ and $\gamma(s,x)=\int_{0}^{x}t^{s-1}e^{-t}dt.$
The characteristic function of random variable $(N-n)\sigma^2(n)$ is a product of characteristic functions of gamma distributions
$$\phi_{(N-n)\sigma^2(n)}(t)=\prod_{j=1}^{N-n+1}\frac{1}{\left[1-2\lambda_j(n)it\right]^{1/2}}.$$ Therefore based on representation \eqref{for4} we get the average value for $\sigma^2(n)$
$$E\left[\sigma^2(n)\right]=\frac{1}{N-n}\sum_{j=1}^{N-n+1}\lambda_j(n)=\frac{1}{N-n}tr\left(\tilde{\Sigma}\right),$$
where $tr\left(A\right)$ is a trace of the matrix $A.$ That gives the same result for the mean of $\sigma^2(n)$ as in \eqref{for3b} and connects eigenvalues $\lambda_j(n)$ with a dependence structure of the observed process $X(j).$ Representation \eqref{for4} provides also the variance formula
$$Var\left[\sigma^2(n)\right]=\frac{1}{(N-n)^2}\sum_{j=1}^{N-n+1}\lambda_j^2(n)Var\left[U_j\right]=\frac{2}{(N-n)^2}\sum_{j=1}^{N-n+1}\lambda_j^2(n)=\frac{2}{(N-n)^2}tr\left(\tilde{\Sigma}^2\right),$$
which is the same as in \eqref{for3d}.\\
The generalized chi-squared distribution in \eqref{for4} was intensively studied. In the literature, there are many different representations for PDF or CDF of such distribution f.e. in terms of zonal polynomials and confluent
hypergeometric functions \cite{Mat82}, single gamma-series \cite{Mos85},
Lauricella multivariate hypergeometric functions \cite{Aal05}, extended
Fox’s functions \cite{Ans14} and others \cite{Vel09}. Here we present the formulas for PDF and CDF according to \cite{Mos85}. The PDF of $\sigma^2(n)$ has a form for $x>0$
\begin{equation}\label{for5}
f_n(x)=C\sum_{k=0}^{\infty}\frac{\delta_kx^{\frac{N-n}{2}+k-1}\exp\left(-\frac{x(N-n)}{2\lambda_1(n)}\right)}{\Gamma\left(\frac{N-n}{2}+k\right)\left(\frac{2\lambda_1(n)}{N-n}\right)^{\frac{N-n}{2}+k}},
\end{equation}
where $\lambda_1(n)$ is the smallest eigenvalue of the matrix $\tilde{\Sigma}$ and
\begin{equation}
C=\prod_{j=1}^{N-n+1}\left(\frac{\lambda_1(n)}{\lambda_j(n)}\right)^{1/2},\quad
\gamma_k=\sum_{j=1}^{N-n+1}\frac{(1-\lambda_1(n)/\lambda_{j}(n))^k}{2k},\quad
\delta_{k+1}=\frac{1}{k+1}\sum_{j=1}^{k+1} j  \gamma_j \delta_{k+1-j}, \quad \delta_0=1.
\end{equation}
The PDF in \eqref{for5} can be understood as a series of densities of gamma distributions $G((N-n)/2+k,2\lambda_1(n)/(N-n)):$
$$f_{n}(x)=C\sum_{k=0}^{\infty}\delta_{k}f_{\left(\frac{N-n}{2}+k,\frac{2\lambda_1(n)}{(N-n)}\right)}(x),\ (x>0).$$
Moreover justified term-by-term integration leads to the CDF formula of $\sigma^2(n)$:
\begin{equation}
F_{n}(x)=P\left(\sigma^2(n)\leq x\right)\label{cdf}
=C\sum_{k=0}^{\infty}\delta_k\int_{0}^{x}{f_{\left(\frac{N-\tau}{2}+k,\frac{2\lambda_1(\tau)}{(N-\tau)}\right)}(y)}dy
=C\sum_{k=0}^{\infty}{\delta_kF_{\left(\frac{N-\tau}{2}+k,\frac{2\lambda_1(\tau)}{(N-\tau)}\right)}(x).}
\end{equation}
We have also the formula for the tail of random variable $\sigma^2(n)$:
\begin{equation}
P\left(\sigma^2(n)> x\right)=1-F_n(x)
=1-C\sum_{k=0}^{\infty}{\delta_kF_{\left(\frac{N-\tau}{2}+k,\frac{2\lambda_1(\tau)}{(N-\tau)}\right)}(x).}
\end{equation}
Therefore we obtain:
\begin{eqnarray*}
P\left(\sigma^2(n)> x\right)=
C\sum_{k=0}^{\infty}\delta_k\Gamma\left(\frac{N-\tau}{2}+k,\frac{x(N-\tau)}{2\lambda_1(\tau)}\right),
\end{eqnarray*}
where $\Gamma(s,x)$ is upper incomplete gamma function defined as
$\Gamma(s,x)=\int_{x}^{\infty}{t^{s-1}e^{-t}dt.}$

\section{Statistical test based on DMA}\label{sec:test}
Knowing the exact probability distribution of the random variable $\sigma^2(n)$ we can propose the statistical test. Because of the generality of this distribution the test is general for any centered Gaussian process. In this paper, we concentrate on the FBM denoted by $B_H(j)$, defined by its covariance function
$$E\left(B_H(j),B_H(k)\right)=D\left(j^{2H}+k^{2H}-|j-k|^{2H}\right),$$
where $D$ is a scale parameter called diffusion constant and $H$ is a self-similarity parameter also called Hurst index.
The null hypothesis of proposed statistical test is 
$$\mathcal{H}_0: \{B_H(1),B_H(2),\ldots,B_H(N)\} \textrm{ is a trajectory of FBM with parameters $D$ and $H$},$$
while alternative hypothesis is
$$\mathcal{H}_1: \{B_H(1),B_H(2),\ldots,B_H(N)\} \textrm{ is not a trajectory of FBM with parameters $D$ and $H$}.$$
The test statistic is the random variable 
$\sigma^2(n)$ distributed according to the CDF of the form \eqref{cdf}.
Therefore we define the $p$-value of the test as the double-tailed event probability
\begin{equation}\label{pvalue}
p=2\min\{P(\sigma^2(n)<t),P(\sigma^2(n)>t)\}=2C\min\left\{\sum_{k=0}^{\infty}\delta_k\Gamma\left(\frac{N-\tau}{2}+k,\frac{t(N-\tau)}{2\lambda_1(\tau)}\right),\sum_{k=0}^{\infty}{\delta_k\frac{\gamma\left(\frac{N-\tau}{2}+k,\frac{t(N-\tau)}{2\lambda_1(\tau)}\right)}{\Gamma\left(\frac{N-\tau}{2}+k\right)},}\right\}
\end{equation}
where $t$ is the value of DMA statistics $\sigma^2(n)$ calculated for empirical trajectory of data. Because $p$-value in \eqref{pvalue} has infinite series representation one has to truncate the sum and compute it as finite truncated sum, where $M$ is the truncation parameter. The error of such approximation was studied in details in \cite{Mos85}. From our perspective it is enough to apply Monte Carlo simulations and compute $p$-value as an empirical quantile from sample of generalized chi-squared random variables of the form $1/(N-n)\sum_{j=1}^{N-n+1}\lambda_j(n)U_j.$\\
Summarizing the procedure for testing hypothesis $$\mathcal{H}_0: \{B_H(1),B_H(2),\ldots,B_H(N)\} \textrm{ is a trajectory of FBM with parameters $D$ and $H$},$$ is the following:
\begin{itemize}
\item[Step 1)]{For empirical trajectory $\{B_H(1),B_H(2),\ldots,B_H(N)\}$ compute DMA statistic 
$$\sigma^2(n)=\frac{1}{N-n}\sum_{j=n}^{N}(X(j)-\tilde{X}_{n}(j))^2:=t.$$}
\item[Step 2)]{Compute the matrix $\tilde{\Sigma}=\{E\left[Y(j)Y(k)\right]:j,k=1,2,\ldots,N-n+1\}$ and its eigenvalues $\{\lambda_j(n):j=1,2,\ldots,N-n+1\}.$}
\item[Step 3)]{$L$ times generate a sample $\mathbf{U}^l=\{U^l_1,U^l_2,\ldots,U^l_{N-n+1}\}$ from chi-squared distribution with one degree of freedom, $l=1,2,\ldots,L.$}
\item[Step 4)]{$L$ times compute the value of generalized chi-squared random variable
$$\sigma^2_l(n)=\frac{1}{N-n}\sum_{j=1}^{N-n+1}\lambda_j(n)U^l_j,\qquad l=1,2,\ldots,L.$$
}
\item[Step 5)]{Compute double-tailed event $p$-value as
$$p=\frac{2\min\left\{\#\{\sigma^2_l(n)>t\},\#\{\sigma^2_l(n)<t\}\right\}}{L}.$$
If $p<\alpha$ reject the null hypothesis $\mathcal{H}_0,$ where $\alpha$ is a significant level. In other case there is no significant statistical proof for rejection of $\mathcal{H}_0.$}
\end{itemize}

We propose the test based on the distribution of DMA statistic $\sigma^2(10)$ for argument $n=10,$ because random variable $\sigma^2(10)$ has different domains for different values of Hurst index $H$ in the case of fixed scale parameter $D.$ However two issues need further studies. The first open problem is the optimization of the proposed test according to the argument $n.$ Natural questions arise about the optimal choice of $n$ and the most effective performance of the test. The crucial point is the dependence of the domain of DMA test statistic on the different values of Hurst exponent $H.$ The second issue is the problem of the testing procedure in the case of unknown scale parameter $D.$ The essence of this problem is that the DMA test statistic can have not disjoint domains for different pairs of parameters $(D,H).$ These problems need continuing research and will be developed by the author. 

In the simulation examination of the proposed statistical test, we consider the case of standard FBM model with fixed $D=1.$

\section{Monte Carlo simulations}\label{sec:sim}
\begin{figure}[h]
	\centering
		\includegraphics{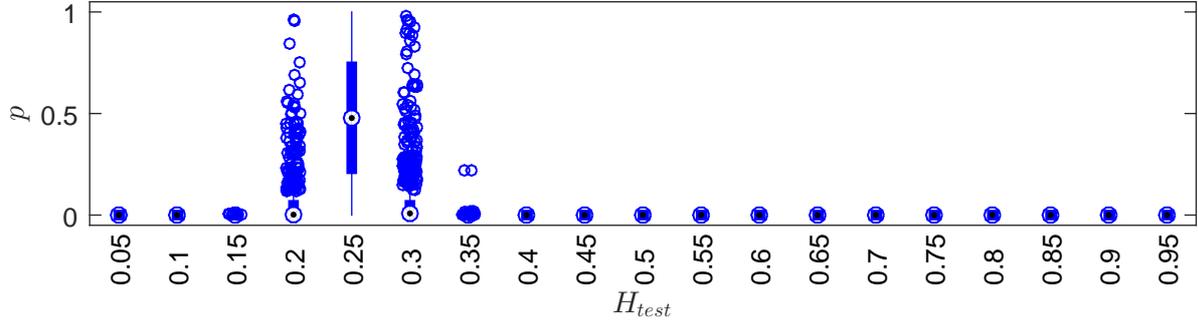}
	\caption{$p$-values obtained from $T=1000$ Monte Carlo simulations from testing $\mathcal{H}_0$ for any $H_{test}\in\{0.05,0.1,\ldots,0.95\}.$ The significant level was $\alpha=0.05$ and the $H_{real}=0.25.$}
	\label{fig1}
\end{figure}

In order to examine the proposed test, we perform Monte Carlo simulations.
First we present results for the case of the Hurst index $H_{real}=0.25,$ which corresponds exemplary subdiffusion case. We generate $T=1000$ independent trajectories of FBM process with fixed $D=1$ and length $N=1000.$ For each trajectory we test the null hypothesis
$$\mathcal{H}_0: \{B_H(1),B_H(2),\ldots,B_H(N)\} \textrm{ is a trajectory of FBM with $H_{test}$},$$ where $H_{test}\in\{0.05,0.1,\ldots,0.95\}.$ Therefore for each case of $H_{test}$ we test $\mathcal{H}_0$ 1000 times and obtain 1000 corresponding $p$-values. On the Figure \ref{fig1} we present boxplots of obtained $p$-values for all cases of $H_{test}.$ The results for $H_{test}<0.2$ and $H_{test}>0.3$ are almost all $p<0.05$ and that is the strong statistical evidence to reject incorrect $\mathcal{H}_0.$ For $H_{test}=0.2$ and $H_{test}=0.3$ we obtained 767 and 751 results with $p<0.05$ respectively. That means more than 75\% of correct rejections of incorrect $\mathcal{H}_0.$ In the case when $H_{test}=0.25$ and $\mathcal{H}_0$ is true we obtained 76 results with $p<0.05.$ In other words we made a type I error (incorrect rejection of true $\mathcal{H}_0$) 7.6\% of $T=1000$ tests. The detailed numbers of accepting of $\mathcal{H}_0$ or $\mathcal{H}_1$ for the case with $H_{real}=0.25$ we present in Table \ref{tab1}. 

\begin{table}[h]
	\centering
		\begin{tabular}{|c|c|c|c|c|c|c|c|c|c|c|}
		\hline
		$H_{test}$ & 0.05 & 0.1 & 0.15 & 0.2 & 0.25 & 0.3 &0.35 & 0.4 & 0.45 &0.5 \\\hline
		$\mathcal{H}_0$ & 0 & 0 & 0 & 233 & 924 & 249 &2 & 0 & 0 &0 \\\hline
$\mathcal{H}_1$ & 1000 & 1000 & 1000 & 767 & 76 & 751 &998 & 1000& 1000 &1000  \\\hline\hline
\multicolumn{2}{|c|}{$H_{test}$} & 0.55 & 0.6 &0.65 & 0.7 & 0.75 &0.8 & 0.85 & 0.9 & 0.95 \\\hline
\multicolumn{2}{|c|}{$\mathcal{H}_0$} & 0 & 0 & 0 & 0 & 0 & 0 &0 & 0 & 0  \\\hline
\multicolumn{2}{|c|}{$\mathcal{H}_1$} & 1000 & 1000 & 1000 & 1000 & 1000 & 1000 &1000 & 1000 & 1000   \\\hline
		\end{tabular}
	\caption{Numbers of accepting of $\mathcal{H}_0$ or $\mathcal{H}_1$ at the significant level $\alpha=0.05$ for the case with $H_{real}=0.25$ obtained from $T=1000$ Monte Carlo simulations.}
	\label{tab1}
\end{table}

\begin{figure}[h]
	\centering
		\includegraphics{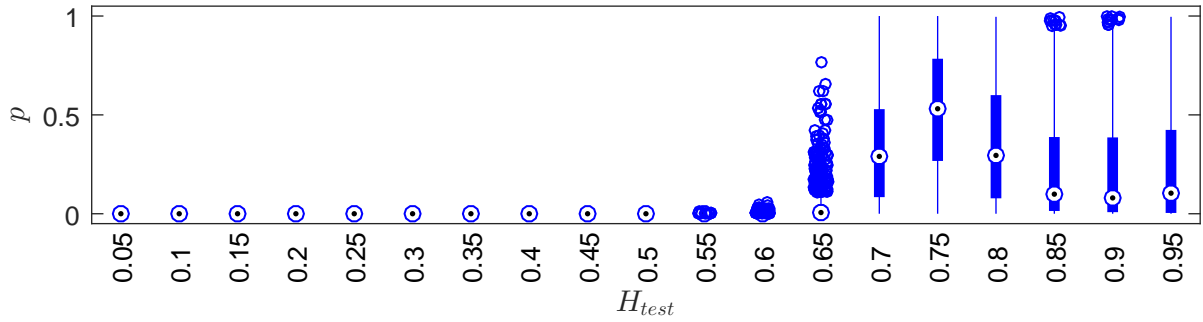}
	\caption{$p$-values obtained from $T=1000$ Monte Carlo simulations from testing $\mathcal{H}_0$ for any $H_{test}\in\{0.05,0.1,\ldots,0.95\}.$ The significant level was $\alpha=0.05$ and $H_{real}=0.75.$}
	\label{fig2}
\end{figure}

The next case is a validation of the proposed test for exemplary superdiffusion case with $H_{real}=0.75.$ Analogous simulations produced 19 sets of $p$-values 
corresponding $H_{test}\in\{0.05,0.1,\ldots,0.95\}.$ The each set contains 1000 $p$-values presented as a boxplot on the Figure \ref{fig2}. The results for
$H_{test}<0.65$ are almost all $p<0.05$ and that is the strong statistical evidence to reject incorrect $\mathcal{H}_0.$ For cases with $H_{test}=0.7$ and $H_{test}>0.75$ the test works not so good as for previous subdiffusion scenario. It incorrectly accepts $\mathcal{H}_1$ more than 80\% times for $H_{test}=0.7$ and $H_{test}=0.8$ and around 60\% times for $H_{test}>0.8.$ So for those cases the type II error is committed very often and the power of the test is weak. On the other hand in the case when $H_{test}=0.75$ and $\mathcal{H}_0$ is true we obtained 63 results with $p<0.05.$ In other words we made a type I error 6.3\% of $T=1000$ tests. The detailed numbers of accepting of $\mathcal{H}_0$ or $\mathcal{H}_1$ for the case with $H_{real}=0.75$ we present in Table \ref{tab2}. 

\begin{table}[h]
	\centering
		\begin{tabular}{|c|c|c|c|c|c|c|c|c|c|c|}
		\hline
		$H_{test}$ & 0.05 & 0.1 & 0.15 & 0.2 & 0.25 & 0.3 &0.35 & 0.4 & 0.45 &0.5 \\\hline
		$\mathcal{H}_0$ & 0 & 0 & 0 & 0 & 0 & 0 &0 & 0 & 0 &0 \\\hline
$\mathcal{H}_1$ & 1000 & 1000 & 1000 & 1000 & 1000 & 1000&1000 & 1000& 1000 &1000  \\\hline\hline
\multicolumn{2}{|c|}{$H_{test}$} & 0.55 & 0.6 &0.65 & 0.7 & 0.75 &0.8 & 0.85 & 0.9 & 0.95 \\\hline
\multicolumn{2}{|c|}{$\mathcal{H}_0$} & 0 & 1 & 233 & 844 & 937 & 835 &611 & 568 & 582  \\\hline
\multicolumn{2}{|c|}{$\mathcal{H}_1$} & 1000 & 999 & 767 & 156 & 63 & 165 &389 & 432 & 420   \\\hline
		\end{tabular}
	\caption{Numbers of accepting of $\mathcal{H}_0$ or $\mathcal{H}_1$ at the significant level $\alpha=0.05$ for the case with $H_{real}=0.75$ obtained from $T=1000$ Monte Carlo simulations.}
	\label{tab2}
\end{table}

\section{Conclusion}\label{sec:con}
In this work, we proposed a new statistical test to identify the FBM model in empirical data. This tool is based on the exact probability distribution of the DMA test statistic $\sigma^2(n)$, which is very sensitive due to the Hurst index $H.$ The proposed procedure is a new original result in the theory of statistical inference of Gaussian processes.

Conducted Monte Carlo simulations indicate that the constructed test works worse in the case of the superdiffusion when $H>1/2.$ In such a scenario, the type II error is very often committed and the power of the test seems to be weaker than for the subdiffusion. This is due to the fact that for the superdiffusion, the domains of the test statistic $\sigma^2(n)$ are close to each other and have joint parts to differing $H$ parameters. This is not the case for subdiffusion where the test works much better. However, it should be strongly emphasized that for both sub and superdiffusion the type I error occurs very rarely and the test correctly accepts the null hypothesis when it is true. In connection with such a test performance and the type II errors, it is possible to modify and optimize the proposed procedure. Namely, a better test performance can be obtained by selecting the argument $n$ of the test statistic $\sigma^2(n).$ It is an interesting issue, worth attention and further research.

Another direction of research on the constructed test is its generalized version due to the unknown parameter of the scale parameter $D.$ In this paper, we assumed a standard FBM with $D=1.$ In the general case with the unknown $D,$ the proposed test should be combined with the pre-estimation of the parameter $D$ by other known methods. However, by applying the theory of the ratios of quadratic Gaussian forms \cite{MatPro92}, it is possible to generalize the described test to the situation of the unknown and non-estimated parameter $D.$ In this case, the probability distribution of the ratios of quadratic forms will not depend on $D$ at all and this parameter will be irrelevant.

The described test procedure can be applied in a sequential manner according to the grid of the values of the parameter $H.$ This will allow to reject the hypotheses with false $H$ values and accept the FBM hypothesis with the true $H.$ Such performance of the proposed test will also be a method for estimating the Hurst exponent as well as a reliable test procedure.

Finally, we want to point out that the statistical test proposed for FBM can be generalized (due to the theory of Gaussian quadratic forms) for each non-degenerated Gaussian process.

\section{Appendix}
\label{sec:app}
Here we present the Matlab code for the proposed statistical test.\\
\ttfamily

\definecolor{mblue}{rgb}{0,0,1} 
\definecolor{mgreen}{rgb}{0.13333,0.5451,0.13333} 
\definecolor{mred}{rgb}{0.62745,0.12549,0.94118} 
\definecolor{mgrey}{rgb}{0.5,0.5,0.5} 
\definecolor{mdarkgrey}{rgb}{0.25,0.25,0.25} 
  
\DefineShortVerb[fontfamily=courier,fontseries=m]{\$} 
\DefineShortVerb[fontfamily=courier,fontseries=b]{\#} 
  
\noindent                                                                
 $$\color{mblue}$function$\color{black}$ [h,p,t]=DMAtest(x,n,H,D,alpha,L)$\\
 $$\color{mgreen}$
 $$\color{mgreen}$
 $$\color{mgreen}$
 $$\color{mgreen}$
 $$\color{mgreen}$
 $$\color{mgreen}$
 $$\color{mgreen}$
 $$\color{mgreen}$
 $$\color{mgreen}$
 $$\color{mgreen}$
 $$\color{mgreen}$
 $$\color{mgreen}$
 $$\color{mgreen}$
 $$\color{mgreen}$
 $$\color{mgreen}$
 $$\color{mgreen}$
 $$\color{mgreen}$
 $$\color{mgreen}$
 $$\color{mgreen}$
 $$\color{mgreen}$
 $$\color{mgreen}$
 $$\\
 $$\color{mgreen}$
 $N=length(x);$\\
 $xmean=sum(x(repmat([1:n]',1,N-n+1)+repmat(0:N-n,n,1)))/n;$\\
 $t=sum((x(n:N)-xmean).^2)/(N-n);$\\
 $$\\
 $$\color{mgreen}$
 $$\color{mgreen}$
 $R=repmat([1:N]',1,N);$\\
 $C=R';$\\
 $X=D*(R.^(2*H)+C.^(2*H)-abs(R-C).^(2*H));$\\
 $$\\
 $$\color{mgreen}$
 $Y1=zeros(N-n+1,N-n+1);$\\
 $Y2=Y1;$\\
 $Y3=Y1;$\\
 $Y=Y1;$\\
 $$\color{mblue}$for$\color{black}$ i=1:N-n+1$\\
 $    $\color{mblue}$for$\color{black}$ j=1:N-n+1$\\
 $        Y1(i,j)=(1-1/n)^2*X(i+n-1,j+n-1);$\\
 $        Y2(i,j)=(1/(n^2)-1/n)*(sum(X(i+n-1,j:j+n-2))+sum(X(i:i+n-2,j+n-1)));$\\
 $        Y3(i,j)=1/(n^2)*sum(sum(X(i:i+n-2,j:j+n-2)));$\\
 $    $\color{mblue}$end$\color{black}$$\\
 $$\color{mblue}$end$\color{black}$$\\
 $Y=Y1+Y2+Y3;$\\
 $lambda=eig(Y)';$\\
 $$\\
 $$\color{mgreen}$
 $U_j=chi2rnd(1,N-n+1,L);$\\
 $$\\
 $$\color{mgreen}$
 $sigma_j=1/(N-n)*lambda*U_j;$\\
 $$\\
 $$\color{mgreen}$
 $p=2*min(sum(sigma_j>t)/L,sum(sigma_j<t)/L);$\\
 $$\color{mblue}$if$\color{black}$ p<alpha$\\
 $    h=1;$\\
 $$\color{mblue}$else$\color{black}$$\\
 $    h=0;$\\
 $$\color{mblue}$end$\color{black}$$\\
 $$\\
 $$\\
 $$\\ 
\UndefineShortVerb{\$} 
\UndefineShortVerb{\#}
\normalfont
\vspace{-2cm}
\section*{Acknowledgements} 
The author would like to acknowledge the support of NCN Maestro Grant No. 2012/06/A/ST1/00258.

\end{document}